\documentclass[aps,prl,twocolumn,showpacs,groupedaddress]{revtex4}
\usepackage{amssymb}

\usepackage[dvips]{graphicx}

\begin{document}

\title{Superconducting phase coherent electron transport in proximity
conical ferromagnets}
\author{I. Sosnin}
\author{H. Cho}
\author{V.T. Petrashov}
\affiliation{Department of Physics, Royal Holloway, University of
London, Egham, Surrey TW20 0EX, UK}

\author{A.F. Volkov}

\affiliation{Theoretische Physik III, Ruhr-Universit\"{a}t Bochum,
D-44780 Bochum, Germany}

\affiliation{Institute of Radioengineering and Electronics of the
Russian Academy of Sciences, 125009 Moscow, Russia }
\date{\today}

\begin{abstract}
We report superconducting phase-periodic conductance oscillations
in ferromagnetic wires with interfaces to conventional
superconductors. The ferromagnetic wires were made of Ho, a
conical ferromagnet. The distance between the interfaces was much
larger than the singlet superconducting penetration depth. We
explain the observed oscillations as due to the long-range
penetration of an unusual "helical" triplet component of the order
parameter that is generated at the superconductor/ferromagnet
interfaces and maintained by the intrinsic rotating magnetization
of Ho.
\end{abstract}


\maketitle


The superconducting proximity effects in normal ($N$) non-magnetic
metals \cite{Lambert} can be described by the three characteristic
length scales: the \textit{coherence length} $\xi_{N}$, the
electron \textit{phase breaking length} $L_{\phi}$, and the
electron \textit{mean free path}, $l$. The normal metal may become
superconducting within $\xi_{N}$ from the normal/superconducting
($N/S$) interfaces, while the superconductor-induced changes in
the normal conductance may survive even longer distances up to
$L_{\phi}$. Here we specialize in proximity effects in thin films,
where $l$ belongs to the nanometer scale due to the elastic
electron scattering. It is usually the smallest relevant length.
The values of $\xi_{N}=\sqrt{\hbar D/k_{B}T}$ and
$L_{\phi}=\sqrt{D\tau_{\phi}}$ may reach the micrometer scale at
low temperatures; $D=v_{F}l/3$ is the electron diffusion
coefficient, $k_{B}$ is Boltzman constant, \textit{T} is the
temperature, $v_{F}$ is the Fermi velocity, and $\tau_{\phi}^{-1}$
is the electron phase breaking rate. Normal proximity conductors
with the dimensions less than $L_{\phi}$ are, in essence, electron
quantum interferometers showing periodic conductance oscillations
as a function of superconducting phase difference between the
$N/S$ interfaces \cite{PetPRL95}.

When the normal metal is substituted with a ferromagnet ($F$), the
superconducting correlations and phase coherent effects are
destroyed by exchange field within the ferromagnetic coherence
length $\xi_{F}$ which is equal to $\sqrt{\hbar D/I}$ when
$I\tau/\hbar<1$ and equal to $l$ when $I\tau/\hbar>1$, where $I$
is the exchange splitting of conduction bands and $\tau$ is the
elastic scattering time \cite{VolkovPRB01, Buzdin}. The reason why
conventional singlet superconducting correlations do not survive
in a ferromagnet beyond $\xi_{F}$ is well understood: the exchange
field causes the splitting of spin-up and spin-down conduction
bands so that a wavefunction of a singlet-state electron pair
acquires a fast oscillating part due to the net momentum of the
pair. The impurity scattering smears these oscillations and leads
to the exponential decay of the condensate function.

\begin{figure}[h]
\label{f:01}
\par
\begin{center}
\includegraphics[angle=0,width=8cm,keepaspectratio,clip]{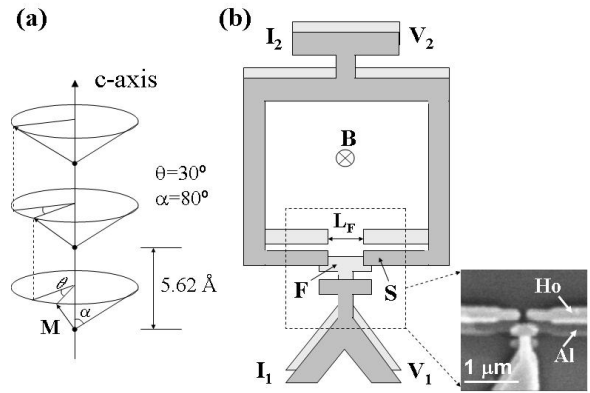}
\end{center}
\caption{a) Magnetic structure of Ho: magnetization \textbf{M}
rotates by 30$^{\circ}$ each atomic layer along $c$-axis at an
angle of 80$^{\circ}$ to this axis. b) Experimental set-up and SEM
micrograph of $S/F/S$ junction area prepared by shadow
evaporation.}
\end{figure}

The possibility of a long-range condensate penetration at
distances much larger than $\xi_{F}$ was not discussed until
recently when new experiments with $F$-wires contacting
superconductors were undertaken \cite{PetJETP94, Giroud,
PetPRL99}. The experiments were of two kinds: those with the
measuring current flowing through the ferromagnetic conductors
bypassing the superconductors and those with the current flowing
through the $F/S$ interfaces. Both could be interpreted as a
manifestation of a long-range superconducting proximity effect.
However, as the experiments strongly disagreed with the existing
theory, efforts were made to explain the results in a way that
didn't involve long-range superconducting correlations. Purely
interfacial phenomena were suggested to explain the measurements
with the current flowing through the $F/S$ interface \cite{Chand,
Nazarov}. As the interfacial phenomena were not able to account
for the experiments with current bypassing the interfaces, an
alternative explanation has been suggested based on long-distance
domain redistribution at the onset of superconductivity
\cite{Geim}. However, the latter effect was shown to be too small
to account for the superconductor-induced changes in the
resistance \cite{Nugent} reviving the question of anomalous
long-range penetration of superconducting correlations into $F$.
Several theories were put forward recently suggesting
fundamentally new long-range proximity effects (see
\cite{VolkovRMP05} and references therein). All of them involved a
\textit{triplet} component of superconductivity that can be
generated near the $F/S$ interfaces. However, the generation of
the triplet component did not necessarily result in the long-range
proximity effect. A successful theory had to provide a mechanism
for survival of the triplet component in the presence of a strong
electron scattering.

\par A mechanism for such a robust triplet proximity effect was
suggested in \cite{VolkovPRL01}. The mechanism is operational in
the presence of a rotating magnetization similar to that in the
Bloch domain walls. To observe the effect experimentally a
rotating magnetization near the interfaces needs to be created.
The effect is predicted to be phase coherent with the
phase-periodic oscillations in the conductance of $F$ wires.
Although there were many publications on transport in $S/F$
structures \cite{PetJETP94, Giroud, PetPRL99, Chand, Nugent}, no
such coherent oscillations have been reported up to date.

In this Letter we report the first measurements of the
superconducting phase-periodic conductance oscillations in
proximity $F$ wires with distance between the $F/S$ interfaces
more than order in magnitude larger than the \textit{singlet}
magnetic coherence length $\xi_{F0}$. We explain the observed
oscillations as a direct result of the long range triplet
proximity effect predicted in \cite{VolkovPRL01}.

\par A rotating magnetization is insured by the use of Ho, a conical
ferromagnet with intrinsic helical magnetic structure shown in
Fig. 1 a \cite{Koehler}. The total magnetic moment of
10.34$\mu_{B}$ ($\mu_{B}$ is the Bohr magneton) in Holmium belongs
to a cone with opening angle $\alpha=80^{\circ}$ and rotates
making a helix along the $c$-axis with turning angle
$\theta=30^{\circ}$ per inter-atomic layer with a net moment of
1.7$\mu_{B}$ per atom. Above 21K the conical ferromagnetic
structure transforms into a spiral antiferromagnetic structure
($\alpha=90^{\circ}$) with Neel temperature of 133K.

The measured structure had the geometry of Andreev interferometer
shown in Fig. 1 b. The $S$ wires were made of Al, a conventional
$s$-wave superconductor. Six samples were fabricated with the
distance, $L_{F}$, between the $F/S$ interfaces equal to 50, 120,
150, 160, 200, and 250 nm. The samples were fabricated using
electron-beam lithography and the shadow evaporation technique.
This method allowed us to make the whole structure without
breaking vacuum thus avoiding the formation of Ho oxide barriers
at the Ho/Al interfaces. The films were thermally deposited at
$6\times10^{-7}$ mbar. First, 40 nm of Ho was evaporated at an
angle -14$^{\circ}$, then 60 nm of Al, at +14$^{\circ}$. Thus, an
Al loop is created with a Ho segment. The area of the
superconducting loop was close to 20$\mu^{2}$. Figure 1 b shows
sample geometry and an SEM micrograph of the $F/S$ part in one of
the measured samples. The resistance of the structure was measured
using standard four probe technique with electrodes connected as
shown in Fig. 1 b, at low frequencies and temperatures between
0.27 and 40 K with a magnetic field up to 2 T applied
perpendicular to the substrate. The resistivity $\rho$ was in the
range 80 - 90 $\mu\Omega$cm for Ho and 0.5 - 0.6 $\mu\Omega$cm for
Al.

\begin{figure}[h]
\label{f:01}
\par
\begin{center}
\includegraphics[angle=0,width=8cm,keepaspectratio,clip]{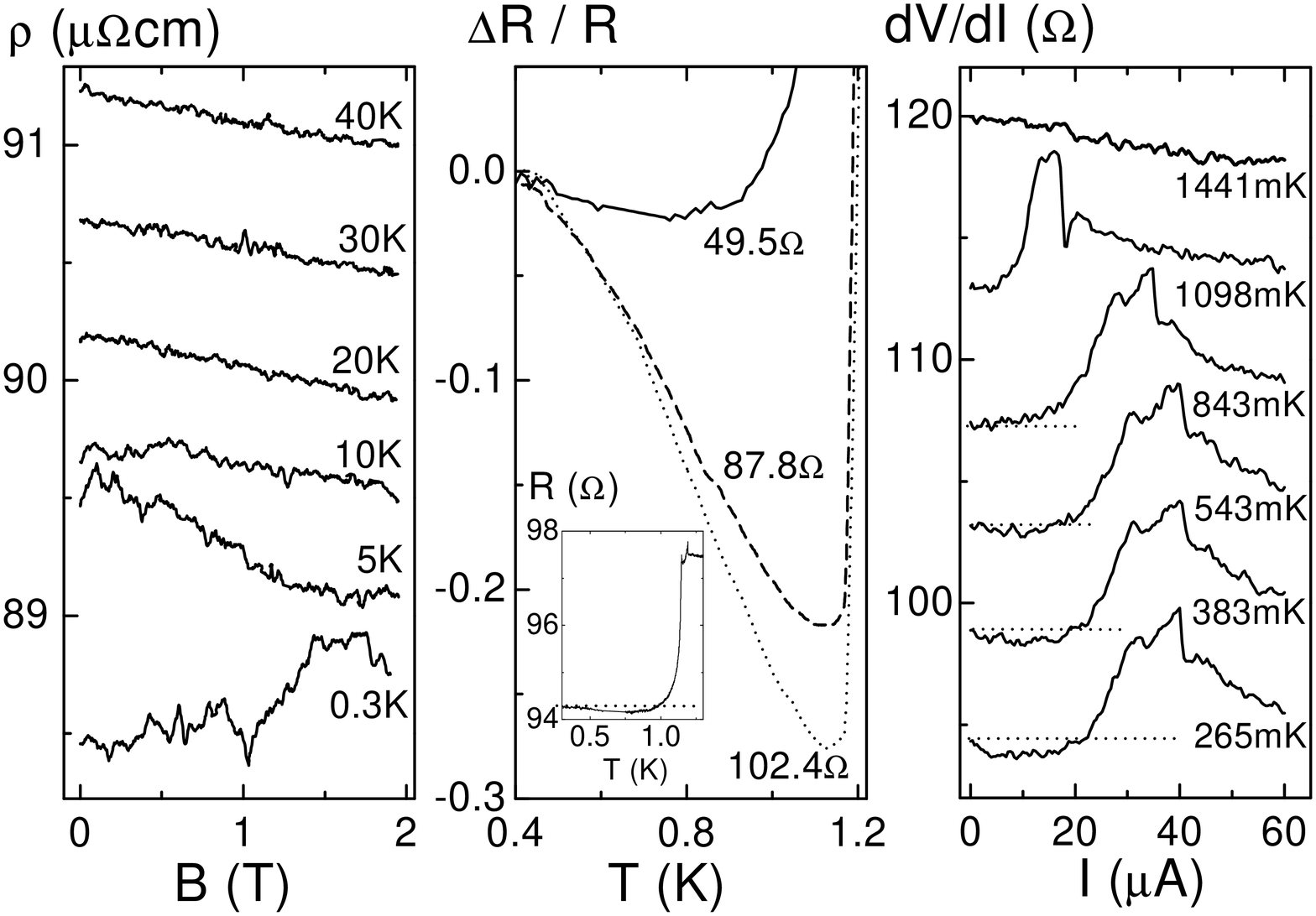}
\end{center}
\caption{a) Magnetoresistance of detached Ho films at different
temperatures. b) Temperature dependence of the resistance of
hybrid Ho/Al structure of Fig. 1 b at $H$ = 0. Inset: Temperature
dependence of the resistance at different Ho/Al interface
resistance. Solid line: $R_{b}=49.5\Omega$; dashed line:
$R_{b}=87.8\Omega$; dotted line: $R_{b}=102.4\Omega$. c)
Differential voltage-current characteristics at different
temperatures in zero magnetic field for sample shown in Fig.1 b.
Curves are offset for clarity.}
\end{figure}

Direct measurements of the magnetoresistance of Ho films
\textit{detached} from superconductors (otherwise deposited in the
same conditions as our hybrid structures) showed ferromagnetic
behavior observed earlier \cite{Singh}. The magnetoresistance
shows typical for Ho change of the sign between 10K and 0.3K (see
Fig. 2 a). The inset in Figure 2 b shows the temperature
dependence of the resistance of our hybrid structure (Fig. 1 b).
The sharp drop at $T$ = 1.2 K corresponds to the superconducting
transition of Al wire and shows that its superconducting
properties are practically not affected by the underlying Ho film.
At lower temperatures the resistance increases. The increase
correlates with the interface resistance as in Ref. [7]. Figure 2
c shows the differential resistance of the hybrid structure at
various temperatures. At low temperatures an increase in the
differential resistance at decreasing voltage is seen, similar to
the increase at lowering temperature of Fig. 2 b. The resistance
increase at low temperatures/voltages is characteristic to the
structures with low transparency interfaces \cite{PetPRL99, Chand,
Nazarov}.

\begin{figure}[h]
\label{f:01}
\par
\begin{center}
\includegraphics[angle=0,width=8cm,keepaspectratio,clip]{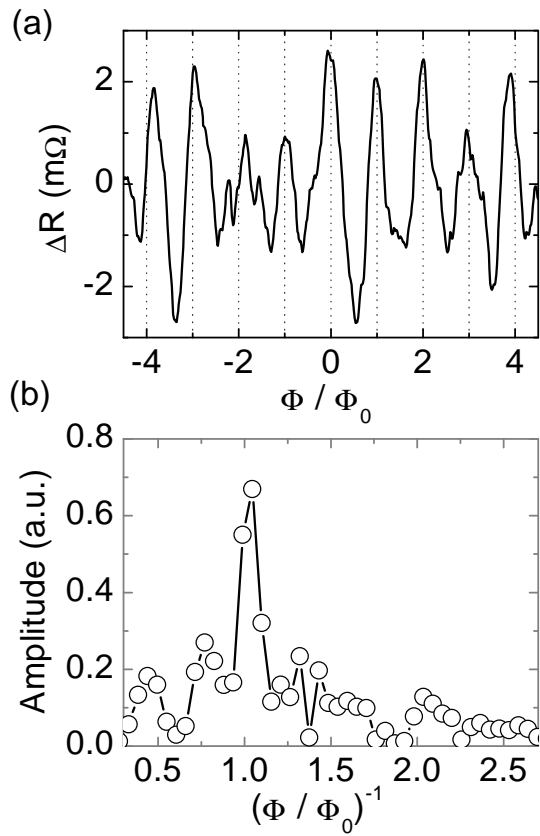}
\end{center}
\caption{a) Magnetoresistance oscillations of the sample shown in
Fig. 1b measured at $T$ = 0.27K as a function of normalized
external flux through the loop. Sample resistance is 94.3$\Omega$.
b) Fourier spectrum of the oscillations confirming the $hc$/2$e$
periodicity.}
\end{figure}

\par The results of measurements of the oscillations in the
resistance of Ho wires as a function of the superconducting phase
difference between their interfaces to superconducting Al are
shown in Figure 3 a. The phase difference was created by magnetic
flux through the $S$ loop with the $F$ segment (Fig. 1 b).  The
period of oscillations corresponded to the flux quantum
$\Phi_{0}=hc/2e=2\times10^{-7}$Gcm$^{2}$ through the area of the
loop with the corresponding sharp peak in the Fourier spectrum of
oscillations (Fig. 3 b). The zero-field resistance of the
structure of  Fig. 3 with Al in superconducting state was $R$ =
94.3 $\Omega$. This resistance can be written as a sum of
contributions from three Ho/Al barriers, $R_{b}$, and that from
ferromagnetic wires, $R_{F}$; $R\approx R_{F}+(3/2)R_{b}$. In our
structures $R_{F}$ is approximately equal to the sheet resistance
that is measured to be 20 $\Omega$, so the relative amplitude of
conductance oscillations is estimated as $\Delta R/R_{F} \approx
10^{-4}$. The oscillations have maximum resistance in zero
magnetic fields. Specially fabricated test $N/S$ structures with
similar geometry and $R_{b}$ values also showed maximum resistance
in zero magnetic field. Out of six measured $F/S$ structures the
oscillations were observed in three samples with $L_{F}$ equal to
50nm, 120 nm and 150 nm. The longer wires of 160, 200nm and 250nm
did not show oscillations within our experimental sensitivity of
$\Delta R/R$ about $10^{-5}$.

To estimate $l$, we use the value of $\rho l \approx
5\times10^{-5}\mu$cm$^{2}$ calculated using data \cite{Ratna,
Reale}. This gives us $l\approx 6$ nm. Another estimate can be
made using the residual resistance ratio ($RRR$) of our Ho films.
The measured $RRR=R(300$K$)/R(0.3$K$)\approx 1.5$ suggesting that
$l$ in our samples is approximately equal to the electron-phonon
mean free path at room temperature that is on the same order as
the above estimate. Using $v_{F}=10^{8}$cm/s we also obtain
electron elastic scattering time $\tau=6\times 10^{-15}$s and
diffusion constant $D=20$ cm$^{2}/$s. We estimate the value of $I$
in Ho using experimental data for the spin polarization, $P$, of
conduction electrons in Ho $P=7\%$ \cite{Meservey}. In the free
electron model we get

\begin{equation}
I=2P\epsilon_{F},
\end{equation}

where $\epsilon_{F}=7.7$ eV is Fermi energy of Ho. Thus, (1)
predicts $I \approx 1.1$ eV. Another estimation of $I$ can be made
using the value of the magnetic moment per atom due to conduction
electrons, $\mu_{CE}$ = 0.34 $\mu_{B}$ \cite{Meservey}, which
gives $I \approx 0.84$ eV in reasonable agreement with the above
estimation. Thus, in our case the parameter $I\tau/\hbar \approx
10$ corresponding to the clean limit, so that $\xi_{F0} \approx
l$.

The observed resistance oscillations prove that the phase
coherence has been established in our ferromagnetic wires through
the length $L_{F}$ up to 150nm. Such a long-range phase coherence
cannot be explained by proximity effect involving penetration of
singlet order parameter. The upper limit for the singlet
penetration length $\xi_{F0}$, as we mentioned before is equal to
$l$. Such short penetration depth rules out singlet proximity
effect as it is attenuated by the factor
exp$(-\xi_{F0}/L_{F})=2\times 10^{-9}$ at $L_{F}$ = 120 nm. This
is in line with the value of $\xi_{F0}=1.2$ nm obtained from
experiments on $SFS$ Josephson junctions with ferromagnetic layer
made of Gd \cite{Bourgeois} another rare-earth element with
ferromagnetism due to localized 4f electrons and indirect exchange
splitting of conduction bands similar to Ho, however, without
ferro-cone magnetization structure.

The "helical" triplet superconductivity contains states with spin
projections $s=\pm1$ that are insensitive to the exchange field,
as all other triplet mechanisms considered recently \cite{Kadig,
Eschrig, Edelstein}. However its condensate function being odd in
the Matsubara frequency (the so called \textit{odd} triplet
superconductivity) is even in the momentum $p$ and and therefore,
unlike other unconventional condensates, is not destroyed by the
presence of non-magnetic impurities thus surviving much longer
distances than the mean free path of quasiparticles. It is
generated in the presence of inhomogeneity of magnetization at the
$F/S$ interfaces. Such inhomogeneities, hence the effect, can in
principle exist in "usual" ferromagnets within the domain walls
explaining experiments \cite{PetJETP94, Giroud, PetPRL99}.
However, no existing technology can create them in a controlled
way at the $F/S$ interfaces with nanoscale precision. In our Ho
conductors the helical magnetization is an intrinsic property. The
value of the helical triplet coherence length, $\xi_{F1}$, depends
on $I$, $\tau$, the period of magnetization rotation in space,
$L_{M}$, and their relationships. The generalized formula for
$\xi_{F1}$ valid for arbitrary relation between $I$ and
$\hbar\tau^{-1}$ can be written as \cite{VolkovUnpubl}:

\begin{equation}
\xi_{F1}^{-1}=\sqrt{2\pi k_{B}T/\hbar D+Q^{2}/(1+(2I\tau/\hbar
)^{2})},
\end{equation}

where $Q=2\pi /L_{M}$, $L_{M}$ is the period of magnetization
rotation. In Ho $L_{M}$ = 6.74 nm. Substituting values for $I$ and
 $\tau$ in (2), we estimate $\xi_{F1} \approx$ 24 nm.

The amplitude of the triplet component $f_{F1}$ of condensate
function can be written as \cite{VolkovUnpubl}

\begin {equation}
f_{F1}(x)=\frac{R_{F}}{R_{b}}\frac{\hbar
Qv_{F}}{2I}exp(-\frac{x}{\xi_{F1}}),
\end{equation}

where $x$ is a space coordinate away from an $F/S$ interface and
$x = 0$ at the interface. The factor exp$(-x/\xi_{F1})$ accounts
for exponential decay of condensate wavefunction over the
characteristic length $\xi_{F1}$ from $F/S$ interface, a good
approximation when $x \gg \xi_{F1}$. We use this form because
$L_{F}>\xi_{F1}$. The amplitude of resistance oscillations depends
on interference of condensate wavefunctions generated at both
interfaces and can be estimated as \cite{VolkovUnpubl}

\begin {equation}
\Delta R =
R_{F}\left(\frac{R_{F}}{R_{b}}\right)^{2}\left(\frac{l}{L_{M}}\frac{\pi
\hbar}{I\tau}\right)^{2}\times exp(-L_{F}/\xi_{F1}),
\end{equation}

Substituting experimental values in (4) we obtain $\Delta R$ = 1.2
m$\Omega$ in a reasonable agreement with experimentally observed
amplitudes (see Fig 3 a).

Note that although in the main approximation the singlet component
penetrates over a small distance $\xi_{F0}$, the coupling of
singlet and triplet order parameters means that there will be a
long-range, of the order $\xi_{F1}$, penetration of singlet
component. However, contribution to resistance from this
long-range part of the singlet component is smaller than the that
of the triplet component by the parameter
$(Qv_{F}l/I\xi_{F1})^{2}$ = 4$\times 10^{-2}$, so it can be
neglected here.

In conclusion, we have observed superconducting phase-periodic
conductance oscillations in ferromagnetic wires with helical
magnetization coupled to a singlet superconductor.  The length of
the wires was much larger than the singlet order parameter
penetration depth ruling out the conventional proximity effect. We
explain the oscillations as due to the long range penetration of
unusual helical triplet component of superconductivity that is
generated in ferromagnetic conductors in the presence of rotating
magnetization.

The work was supported by EPSRC grant ARF/001343 (UK) and by SFB
grant 491(Germany). VTP and IS thank V. Edelstein for fruitful
discussions.


\begin{thebibliography}{99}
\bibitem{Lambert} C.J. Lambert and R. Raimondi, J. Phys.: Condens. Matt. \textbf{%
10}, 901 (1998).

\bibitem{PetPRL95} V.T. Petrashov, V.N. Antonov, P. Delsing, and T.
Claeson, Phys. Rev. Lett., \textbf{74}, 5268 (1995).

\bibitem{VolkovPRB01} F.S. Bergeret, A.F. Volkov and K.B. Efetov, Phys. Rev.
B \textbf{64}, 134506 (2001).

\bibitem{Buzdin} I. Baladie and A. Buzdin, Phys. Rev. B
\textbf{64}, 224514 (2001).

\bibitem{PetJETP94}  V.T. Petrashov, V.N. Antonov, S.V. Maksimov, R.S.
Shaikhaidarov, JETP Lett. \textbf{59},523 (1994); V.T. Petrashov, Czech. J.
Phys. \textbf{46}, 3303 (1996).

\bibitem{Giroud}  M. Giroud, H. Courtois, K. Hasselbach, D. Mailly, and B.
Pannetier, Phys. Rev. B \textbf{58}, R11872 (1998).

\bibitem{PetPRL99}  V.T. Petrashov, I.A. Sosnin, I. Cox, A. Parsons, and C.
Troadec, Phys. Rev. Lett. \textbf{83}, 3281 (1999).

\bibitem{Chand}  J. Aumentado and V. Chandrasekhar, Phys. Rev. B \textbf{64},
54505 (2001).

\bibitem{Nazarov}  W. Belzig, A. Brataas, Y.V. Nazarov, and G.E.W. Bauer, Phys.
Rev. B \textbf{62}, 9726 (2000).

\bibitem{Geim}  S.V. Dubonos, A.K. Geim, K.S. Novoselov, and I.V. Grigorieva,
Phys. Rev. B \textbf{65}, 220513(R) (2002).

\bibitem{Nugent}  P. Nugent, I. Sosnin and V.T. Petrashov, J. Phys.: Condens.
Matt. \textbf{16}, L509 (2004).


\bibitem{VolkovRMP05} F.S.Bergeret, A.F.Volkov and K.B.Efetov, Rev.Mod.Phys. \textbf{77}, 4 (2005).

\bibitem{VolkovPRL01}  F.S. Bergeret, A.F. Volkov and K.B. Efetov, Phys. Rev. Lett.
\textbf{86}, 4096 (2001).


\bibitem{Singh}  R.L. Singh, Phys. Rev. B, \textbf{15} 4174 (1977).

\bibitem{Koehler}  W.C. Koehler, J.W. Cable, M.K. Wilkinson, and E.O. Wollan,
Phys. Rev. \textbf{151}, 414 (1966).

\bibitem{Meservey}  R. Meservey, D. Paraskevopoulos, and P.M. Tedrow, Phys. Rev.
B, \textbf{22}, 1331 (1980).

\bibitem{Bourgeois}  O. Bourgeois, P. Gandit, J. Lesuer, A. Sulpice, X. Grison, and
J. Chaussy, Eur. Phys. J. B \textbf{21}, 75 (2001).

\bibitem{Ratna}  R. Ratnalingam and J.B. Sousa, Phys. Lett., \textbf{30A}, 8
(1969).

\bibitem{Reale}  C. Reale, Appl. Phys., \textbf{2}, 183 (1973).

\bibitem{Kadig}  A. Kadigrobov, R. I. Shekhter and M. jonson, Europhys. Lett.
\textbf{54}, 394 (2001).

\bibitem{Eschrig}  M. Eschrig, J. Kopu, J. C. Cuevas, and G. Sch\"{o}n, Phys.
Rev. Lett. \textbf{90}, 137003 (2003).

\bibitem{Edelstein}  V.M. Edelstein, Phys. Rev. B \textbf{67},
020505(R) (2003).

\bibitem{VolkovUnpubl}  A.F. Volkov, unpublished.


\end{thebibliography}

\end{document}